\documentclass[12pt,onecolumn]{IEEEtran}

\usepackage{amssymb}
\usepackage{amsmath}
\usepackage{graphicx}
\usepackage{cite}
\usepackage{balance}
\usepackage[utf8x]{inputenc}
\usepackage{fancyhdr}
\usepackage{amsmath,amssymb,amsthm}
\usepackage{graphicx,epstopdf}
\usepackage{epsfig}	
\usepackage{subfigure}
\usepackage{lipsum}
\usepackage{subfigure}
\usepackage[algo2e,ruled, vlined, linesnumbered]{algorithm2e}
\usepackage{bbm}
\usepackage{stfloats}
\usepackage{algorithm}
\usepackage{algorithmic}
\usepackage{bm}
\usepackage{lettrine}

\newtheorem{theorem}{Theorem}

\newtheorem{lemma}{Lemma}

\newtheorem{corollary}{Corollary}

\newtheorem{definition}{Definition}
\renewcommand{\baselinestretch}{1.75}

\makeatletter
\def\ScaleIfNeeded{%
	\ifdim\Gin@nat@width>\linewidth \linewidth \else \Gin@nat@width
	\fi } \makeatother

\begin{document}
	
	\title{Self-Controlled Jamming Resilient Design Using Physical Layer Secret Keys}
	\author{Jay~Prakash, ~\IEEEmembership{Student Member,~IEEE,} Chenxi~Liu,~\IEEEmembership{Member,~IEEE,} Jemin~Lee,~\IEEEmembership{Member,~IEEE,} and Tony~Q.~S.~Quek,~\IEEEmembership{Fellow,~IEEE}
		\thanks{J. Prakash, C. Liu, and T. Q. S. Quek are with the Information Systems Technology and Design Pillar, Singapore University of Technology and Design,
			Singapore 487372 (email: jay\_prakash@mymail.sutd.edu.sg, chenxi\_liu@sutd.edu.sg, tonyquek@sutd.edu.sg). }
		\thanks{J. Lee is with the Department of Information and Communication Engineering,
			Daegu Gyeongbuk Insitute of Science and Technology (DGIST), Daegu,
			Korea (email: jmnlee@dgist.ac.kr).}
		\thanks{The corresponding author is C. Liu.}
		
	}
	
	\maketitle
	
	\markboth{Submitted to IEEE Transactions on Information Forensics and Security}%
	{Shell \MakeLowercase{\textit{et al.}}: Bare Demo of IEEEtran.cls for IEEE Journals}

	\begin{abstract}
		Direct-sequence spread spectrum (DSSS) has been recognized as an effective jamming resilient technique. However, the effectiveness of DSSS relies on the use of either pre-shared unique secret keys or a bank of public codes, which can be prohibitively expensive in future large-scale decentralized wireless networks, e.g., the Internet of Things. To tackle this problem, in this work we develop a new framework for self-controlled physical-layer-security-based spreading sequence generation. Specifically, we exploit the shared randomness inherent in wireless channels to generate and refresh secret seeds at each communicating node using shared randomness extraction, entropy pooling and random seed generation. The generated secret seeds are then utilized to perform DSSS. To evaluate the performance, we implement our framework on software defined radio platform and examine the successful transmission probability of the system under various models of broadband jamming along with an special case wherein adversary is assumed to have leaked information on key rate. Both our analysis and real-world measurements confirm that communication systems based on our framework can achieve jamming-resilient communications without requiring pre-shared sequences.

	\end{abstract}

	\section{Introduction}
	The Internet of Things (IoT) is a promising aspect of wireless communications, which is envisioned to interconnect ultra-large number of devices from various environments. However, the inherent decentralized structure of the IoT makes devices increasingly vulnerable to potential attacks, as well as making it extremely difficult for traditional centralized security methods to protect devices of the IoT from such attacks. Consequently, developing novel secure transmission schemes that are suitable for large-scale decentralized wireless networks has been receiving more and more research attention.
	
	On the other hand, jamming attacks, among all potential attacks, have been considered as a significant threat to wireless communications in the past few decades, since it can lead to new signal creation, annihilation and/or symbol flipping \cite{jam_nils}. In addition, jamming attacks decrease the signal-to-interference-plus-noise-ratio (SINR) of the legitimate system, thereby directly degrading the bit error rate (BER) and throughput of the legitimate system. Effect of jamming attack on state of art cyber-physical-system (CPS) testbed, SWAT, was studied in \cite{swat} which shows aberrations in physical water cleaning processes, among other effects, in response to jamming of network at different levels in link hierarchy. These detrimental effects of jamming and advancement in narrow-band jamming, as introduced in optimal attack strategy for given modulation in \cite{optijam}, and real-time reactive jammer, as presented in \cite{reactive}, make it imperative to explore jamming resilient protocol.
	
	Traditionally, jamming resilient communications in wireless networks are achieved through spread spectrum techniques such as frequency hopping spread spectrum (FHSS) and direct sequence spread spectrum (DSSS), relying on the assumption that the frequency band used in FHSS and codes used to spread data in DSSS are not known at the jammer. This assumption, however, is less likely to be valid with the low computational complexity of breaking the pre-shared codes in DSSS as well as the rapid increase in the adversary's computational capability.

	To tackle this problem, significant research efforts have been devoted to design jamming resilient communications systems without requiring pre-shared keys \cite{keyless,uncord_key,uncord_fh,pp,pooper,liu_random_dsss,cdma_jamming}. Specifically, in \cite{keyless} and in \cite{uncord_key}, an uncoordinated key agreement protocol was proposed to combat jamming attacks. The network secrecy was examined in \cite{uncord_fh} considering the use of the uncoordinated hopping. Relaxing the requirement on pre-shared secret sequences, the uncoordinated FHSS was proposed in \cite{pp}. Instead of having the same spreading codes for each transmission, in \cite{pooper} the randomized DSSS, which uses a set of codes and randomly choose one of these codes for each transmission, was proposed for peer-to-peer communication systems. As such, an extra processing delay is introduced at the jammer since it needs to decipher the sequence used in each transmission, and thereby improving the jamming resilience of the legitimate system. An improved version of randomized DSSS in \cite{pooper} was later generalized to broadcast communication systems \cite{liu_random_dsss}. In \cite{cdma_jamming}, it was proposed  to use encryption on spread sequence. The aforementioned works \cite{keyless,uncord_key,uncord_fh,pp,pooper,liu_random_dsss,cdma_jamming} focus on the jamming resilient design in centralized communication systems, but the jamming resilient design in decentralized communication systems particularly for the IoT network, where many peer-to-peer communications can occur simultaneously, has not been studied extensively. In such system, significant challenges are introduced for the generation, distribution, and management of spread sequences. In this work, we exploit the potential benefits of physical layer key in enhancing the jamming resilience of decentralized wireless networks. Inspired by the physical layer secret key extraction (e.g., \cite{phyk,sec_key}),  we exploit the shared randomness inherent in wireless channels to perform pseudo-random seed generation. The generated pseudo-random seeds are then used to simultaneously generate spreading codes at all participating nodes of the considered system.
	We note that physical layer key based secure spectrum spread has recently been presented in \cite{sim} and \cite{peter}. A fault tolerant key extraction was proposed in \cite{sim} and the extracted keys were then used directly to perform spectrum spread. A detailed theoretical analysis for physical layer secret key based FHSS was performed in \cite{peter}. These works do not consider two important aspects: functionality assurance in low key rate conditions and orthogonality of spreading sequences thus generated. Compared to \cite{sim} and \cite{peter}, we adopt a cross-layer strategy to use the generated secret keys as the seeds of random sequence generation, and adopt the bit error probability and successful transmission probability as performance metrics. Our contributions are summarized as follows:
	\begin{itemize}
		\item We extract secret keys from orthogonal frequency-division multiplexing (OFDM) subcarriers. In addition, we show how secret key agreement can be achieved in the legitimate system.
		\item
		We propose to utilize cross-layer security techniques for random seed generation. We highlight that our proposed method can support the system requirements even when the key rate is relatively low.
		
		\item To evaluate the performance, we analyze the successful transmission probability of the legitimate system, under active attacks from the jammer. Moreover, we implement our proposed design on software defined radios platform using USRP-2952 and Labview Communication Suite, confirming that our analysis accurately match with real-world measurements.
	\end{itemize}
	
	The rest of the paper is organized as follows. In Section \ref{sec:system_model}, we describe the system model. In Section \ref{sec:randomness_ex}, we detail the shared randomness extraction of the proposed design. In Section \ref{sec:phy_dsss}, we present the components of proposed jamming resilient design and show how the extracted secret keys can be used for secure spectrum spread. We also analyze the performance achieved by the proposed scheme. Numerical
	results and related discussions are presented in Section V.
	Finally, Section VI draws conclusions.
	
	
	{\em Notations}: Column vectors (matrices) are denoted by boldface lower (upper) case letters. Conjugate transpose is denoted by $\left(\cdot\right)^H$. Complex Gaussian distribution is denoted by $\mathcal{CN}$. Statistical expectation is denoted by $\mathbb{E}$. The absolute value of a scalar is denoted by $|\cdot|$. The Frobenius norm of a vector or a matrix is denoted by $\|\cdot\|$.

	\section{System Model}\label{sec:system_model}
	
	\begin{figure}[t]
		\centering
		\subfigure[Channel probing and trasmission.]
		{\includegraphics[width=0.7\columnwidth]{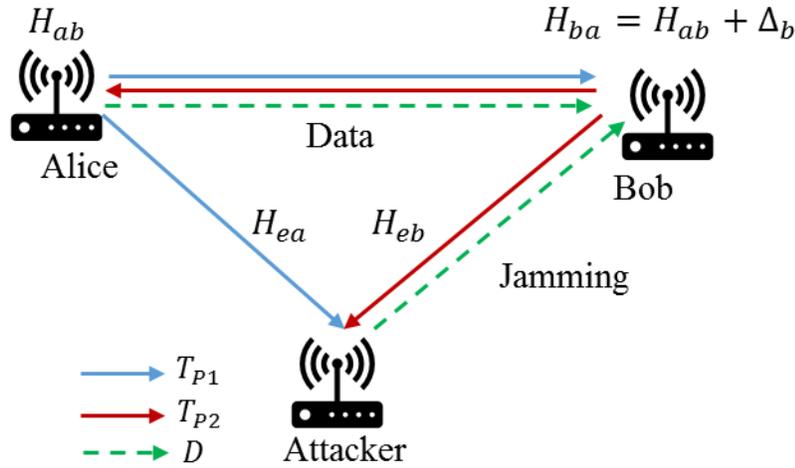}
			\label{proto_a}
		}
		
		\subfigure[TDD frame structure.]
		{	\includegraphics[width=0.7\columnwidth]{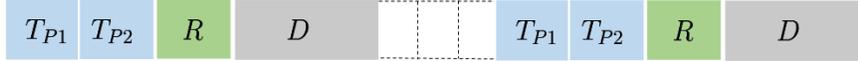}
			\label{prot_b}
		}
		\caption{Illustration of the channel probing and the frame structure of TDD.}
	\end{figure}
	
	
	We consider a wiretap channel, as depicted in Fig.\ref{proto_a}, where a transmitter (Alice) communicates with a legitimate receiver (Bob) in the presence of an {\em active} attacker who can perform malicious jamming attack when legitimate communications occur. We assume that all the nodes are equipped with a single antenna and works in a half-duplex mode. In the considered system, the transmission between Alice and Bob follows a time division duplex (TDD) protocol with the frame slotted structure, as shown in Fig. \ref{prot_b}. Specifically, $T_{P1}$ and $T_{P2}$ denote the channel probing slots when Alice and Bob send each other OFDM based probing symbols. The $m^{\text{th}}$ sample in OFDM symbol transmitted in $T_{P1}$ and $T_{P2}$ can be expressed as
	\begin{align}
	\label{m_sample} x\left[m\right]=\sqrt{M}\sum_{k=0}^{M-1}X[k]\exp\left(j2\pi km/M\right),
	\end{align}
	where $M$ denotes the total number of subcarriers and $X[k]$ denotes the pilot symbol modulated to $k^{\text{th}}$ subcarrier. We assume that the total time length of $T_{P1}$ and $T_{P2}$ is less than the channel coherence time such that the channel reciprocity is guaranteed. In $R$, Alice and Bob generate secret bits independently and establish the common key. In $D$, Alice uses the secret bits to spread the symbols using DSSS.
	%
	\subsection{Channel Model}
	In this work, we consider the frequency-selective broadband channel model. We denote $h_{ij}$, $i,j\in\left\{a,b,e\right\}$ as the channel between node $i$ and $j$, where $a$, $b$, and $e$ stand for Alice, Bob, and attacker, respectively. As such, the impulse response of the channel between node $i$ and $j$ is expressed as
	\begin{align}
	\label{channel_response} h_{ij}\left(\tau,t\right) =\sum_{l=0}^{L-1}h_{ij}[l]= \sum_{l=0}^{L-1} h_{ij}(\tau_l,t)\delta(\tau-\tau_l),
	\end{align}
	where $L$ denotes the total number of channel taps, $\tau_l = lT$ and $h_{ij}\left(\tau_l,t\right)$ denote the delay and attenuation of the $l^{\text{th}}$ channel tap, respectively, $T$ denotes the sampling period, and $\delta(\cdot)$ denotes the Dirac delta function \cite{goldsmith2005wireless}.  Based on \eqref{channel_response}, the frequency response of $h_{ij}\left(\tau,t\right)$ corresponding to the $k^{\text{th}}$ subcarrier can be expressed as
	\begin{align}
	H_{ij}[k]= \sum_{l=0}^{L-1}h_{ij}[l]\exp\left(-{j2\pi kl/M}\right).
	\end{align}
	Without loss of generality, in the following, we omit the subcarrier index $k$.
	
	We then express the $1\times M$ channel vector observed at Bob in the frequency domain as
	\begin{align}
	\label{channel_b} H_{ba}= H_{ab} + \Delta_b,
	\end{align}
	where $\Delta_b$ denotes channel probing errors between Alice and Bob. We also express the $1\times M$ channel vector observed at attacker in the frequency domain as
	\begin{align}
	\label{channel_e} H_{ea}= H_{ab} + \Delta_{e1},
	\end{align}
	where $\Delta_e$ denotes the difference between $H_{ab}$ and $H_{ae}$. We note that the variance of $\Delta_e$ is much larger than that of $\Delta_b$ due to different locations of Bob and attacker.

	%
	
	\subsection{Adversary Model}
	We consider that attacker is of the capability of fast spectrum sensing and RF chain switching. As such, attacker transmits malicious jamming signals only when it senses that the legitimate communication occurs. As previously described, Alice transmits the signal using DSSS. As such, we express the signal received at Bob in the time domain as
	\begin{align}
	\label{received_b} y_b = \sqrt{P_ad_{ab}^{-\alpha}}h_{ab} xC_{ab} +\sqrt{P_ed_{eb}^{-\alpha}}h_{eb}x_e+n_b,
	\end{align}
	where $P_a$ and $P_e$ denote the transmit power at Alice and attacker, respectively, $d_{ab}$ and $d_{eb}$denote the distance between Alice and Bob and the distance between attacker and Bob, respectively, $\alpha$ denotes the path-loss exponent, $C_{ab}$ denotes the spreading code, of length $ L$ bits, used by Alice, $x$ and $x_e$ denotes the transmitted symbol from Alice and the jamming signal from attacker, respectively, and $n_b$ denotes the thermal noise at Bob, which is assumed to be an independent and identically distributed (i.i.d) complex Gaussian random variable with zero mean and variance $\sigma_{b}^2$, i.e., $n_b\sim\mathcal{CN}\left(0,\sigma_b^2\right)$. In \eqref{received_b}, we have $\mathbb{E}\left(|x|^2\right) = 1$ and $\mathbb{E}\left(|x_{e}|^2\right) = 1$.
	
	As per the rule of DSSS, for $C_{ab}$ of length $L$, the signal-to-interference-plus-noise ratio (SINR) at Bob is expressed as \cite{viterbi1995cdma}
	\begin{align}
	\label{SNR_b} \gamma_b = \frac{\gamma_{ab}L|h_{ab}|^2}{\gamma_{eb}|h_{eb}|^2+1},
	\end{align}
	where $\gamma_{ab} = P_ad_{ab}^{-\alpha}/\sigma_b^2$ and $\gamma_{eb} = P_ed_{eb}^{-\alpha}/\sigma_b^2$.

	\section{Shared Randomness Extraction}\label{sec:randomness_ex}
	In this section, we describe in detail how the shared randomness extraction (SRE) (i.e., the physical layer secret key generation) is achieved, which will be used in our proposed scheme. We note that secret key generation is performed by using channel measurements, including the channel impulse response (CIR), the channel frequency response (CFR), and the received signal strength (RSS), as the source of randomness.
	The key benefits of using channel measurements as the source of randomness are:
	\begin{itemize}
		\item \textit{Shared randomness:} Due to the reciprocity of wireless channels, channel measurements are almost same at the transmitter and the receiver.
		\item \textit{Time variant: } Channel measurements are independent of each other across time.
		\item \textit{Position independence:} Channel measurements at two different locations are independent \cite{5fast_key}.
		
	\end{itemize}
	These benefits also stand for primary requirements on the part of source for designing a shared key based secure system.
	
	We note that secret key generation based on different channel measurements has its own suitable scenarios. For example, the RSS-based secret key generation can be cost-effective in systems with low-capability devices, while CIR-based secret key generation is more suitable for the system requiring a higher secret generation rate since CIR-based scheme can generate more keys from multi-path rich environments.
	
	The process of SRE can be divided into three steps (see Fig.~\ref{block} for the block diagram of SRE):
	1)  randomness observation, 2) fuzzy extraction, and 3) privacy amplification,
	which are described in the following subsections.
	
	
	
	
	\subsection{{Randomness Observation: Channel Estimation}}
	For the channel estimation at Alice and Bob, as described in Section II, Alice and Bob exchange OFDM-based pilot probes to estimate channel between them.
	In order to observe highly correlated estimates at both nodes, the following constraint needs to be satisfied:
	\begin{equation}
	{T}_{P1} + T_{P2} + T_{s} \leq T_{c},
	\end{equation} where $T_{s}$ denotes the time spent in switching RF chain at transceivers, and $T_{c}$ denotes the coherence time. We have $T_{c} = \frac{9}{16\pi f_d}$ \cite{5fast_key}, where $f_d$ denotes the Doppler frequency.
	

	
	After channel probing, two highly correlated random matrices $H_{ab}$ and $H_{ba}$ are obtained at Alice and Bob, respectively. We note that both $H_{ab}$ and $H_{ba}$ are arrays of channel measurements across $N$ probing slots and $M$ subcarriers, i.e., $ H_{ba}=[ H_{ba}(f_1),\cdots,H_{ba}(f_M)]$ and $ H_{ab}=[ H_{ab}(f_1),\cdots,H_{ab}(f_M)]$. We also note that $N$ probing slots ($T_{P1}$ and $T_{P2}$) are needed in order to make sure that both nodes have sufficient entropy to support the proposed PHY-DSSS. The value of $N$ is determined by system requirements such as the secret key rate. We also note that the duration between probing slots plays a vital role in generating secret keys with enough entropy. This is because a high probing rate leads to a decreased entropy of generated secret keys, due to high bits correlation across time. As such, we choose the probing rate to be $1/T_{c}$ as in \cite{zhang2014secure} and \cite{ofdm}.
	
	\subsection{Fuzzy Extraction}\label{fuzzy_extraction}
	
	After obtaining these channel estimates at Alice and Bob, the next step is to obtain the same secret keys at Alice and Bob. This can be achieved by the shared randomness extractor \cite{def}, defined as:
	\begin{definition}\label{def} A function $f_{sre}(\cdot)$ is a shared randomness extractor if (1) $f_{sre}(\cdot)$ is a mapping $\{Re\}^n \to \{0,1\}^m $, and $f_{sre}\left(H\right)$, $H\in\left\{H_{ab},H_{ba}\right\}$  is $\epsilon$-close to $U_m$, i.e., $\| <f_{sre}(H,\phi),\phi>- <U_m,\phi>\| \leq \epsilon $, where $U_m$ is the uniform distribution over $\{0,1\}^m$ and $\phi$ is a random variable uniformly distributed over $R$; (2) $f_{sre}\left(H_{ab}\right)$ and $f_{sre}\left(H_{ba}\right)$ return same outputs with at least $1-\epsilon_c$ probability. \end{definition}
	
\begin{figure}[t]
		\centering
		\includegraphics[width=0.7\columnwidth]{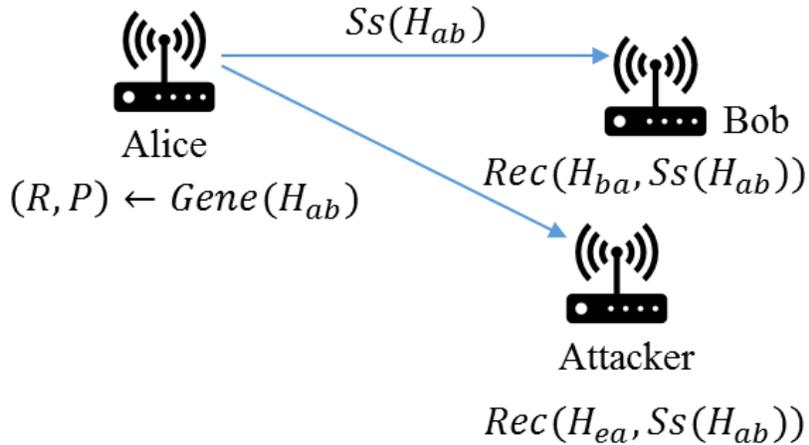}
		\caption{Fuzzy extraction in the presence of attacker.}
		\label{fuzzygen}
	\end{figure}

	One such promising extractor is the \textit{fuzzy extractor}, which can generate sufficiently random bits using two similar and noisy data as inputs \cite{dodis}. Specifically, we define $m$ as the length of the channel estimates $H_{ab}$ at Alice and the channel estimates $H_{ba}$ at Bob, $l$ as the length of output random string $R$, and  $t$ as the maximum distance (e.g., hamming distance or set-difference) between channel estimates at Alice and Bob. How the fuzzy extractor works can be explained as follows:
	fuzzy extractor ${{f (H_{ab},H_{ba},m,l,t)}}$, as shown in Fig. \ref{fuzzygen}, consists of two components: a probability difference function $\textit{Gene}(\cdot)$ and a recovery process $Rec(\cdot)$. First, Alice generates a random string $R \in\{0,1\}^l$ and a helper string  $P\in\{0,1\}^*$ by using $\textit{Gene}(H_{ab})$. In order to obtain the same random string $R$ at Bob, Alice shares the helper string $P$ with Bob. Then, Bob performs the recovery process $Rec\left(H_{ba},P\right)$. If the distance between $H_{ab}$ and $H_{ba}$ is within a certain range, i.e., $dis\left(H_{ab},H_{ba}\right)\leq t$, Bob can perfectly recover the random string $R$. The Fuzzy Extractor also insures that $R$ is $\epsilon$ close to $U_l$, indicating that the generated secret keys are random enough. We highlight that attacker cannot recover the random string $R$ even if it knows $Rec(\cdot)$, due to the fact that $H_{ba}$ and $H_{ea}$ are independent.
	
	We note that there are two types of fuzzy extractors, namely, set-difference based fuzzy extractor and hamming distance based fuzzy extractor. Fig. \ref{fuzzy} shows the block diagram of these two type of fuzzy extractors. We can see that set-difference based fuzzy extractor uses \textit{Pin-sketch} based constructor to design the helper string, while hamming distance based fuzzy extractor utilizes \textit{Secure-sketch} based constructor \cite{dodis}. In this work, we focus on the use of hamming distance based extractor. We will show in Section \ref{result_channel} why the set-difference based fuzzy extractor is not suitable for our considered system. The process of hamming distance based fuzzy extractor can be explained as follows:
	
	\begin{figure}[!t]
		\centering
		\includegraphics[width = 4.5in]{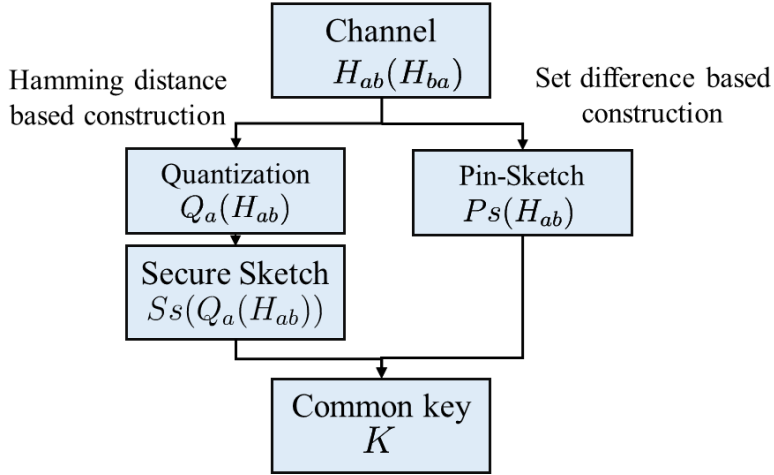}
		\caption{Notions of fuzzy-extraction.}
		\label{fuzzy}
	\end{figure}


	
	\paragraph{{Quantization}}
	We note that Hamming distance based constructor needs binary inputs hence channel observations need to be quantized. Specifically, Alice and Bob use an adaptive quantizer, $\mathbf{Q}_u(\cdot)$, to quantize $H_{ab}$ and $H_{ba}$, respectively. The lower and upper thresholds of $Q_u(\cdot)$ are given by
	\begin{align}
	\mathbf{Q}_u^{+} = \bm{\mu}_{H_{ab}} + \alpha*\bm{\sigma}^2_{H_{ab}},
	\end{align}
	and
	\begin{align}
	\mathbf{Q}_u^{-}  = \bm{\mu}_{H_{ab}} - \alpha*\bm{\sigma}^2_{H_{ab}},
	\end{align}
	respectively, where $\bm{\mu}_{H_{ab}}$ is the mean of channel estimates vector, $\bm{\sigma}$ is the corresponding variance vector, and $\alpha$ denotes a tuning factor. We note that incorporating this tuning factor $\alpha$ in $\mathbf{Q}_{u}(\cdot)$ can decrease secret bit mismatches per block. We drop channel estimates whose amplitudes are between $\textbf{Q}_u^{+}$ and $\textbf{Q}_u^{-}$, while setting values of channel estimates whose amplitudes are above $\textbf{Q}_u^{+}$ and below $\textbf{Q}_u^{-}$ to be $1$ and $0$, respectively.
	
	
	\paragraph{{{Information Reconciliation}}}
	We note that bit mismatches after the quantization are inevitable due to noise, interference, hardware impairments, automatic gain control issues, synchronization issues at Alice and bob, and other issues associated with half-duplex transceivers. Therefore, the next step of \textit{SRE} is to insure both nodes agree on the exact same bit sequence with sufficient randomness.  
	
	We denote the quantized channel estimates at Alice and Bob as $Q_u(H_{ab})$ and $Q_u(H_{ba})$, respectively. Using $Q_u(H_{ab})$, Alice first broadcasts to Bob a \textit{Secure-sketch}, which is given by
	\begin{align}
	SS(Q_u(H_{ab}))= Q_u(H_{ab})-C= Q_u(H_{ab})\oplus C,
	\label{code}
	\end{align}
	where $C$ is a set of $M$ Bose-Chaudhuri-Hocquenghem (BCH)-codes \cite{bch} of $(N',k,2t+1)$ family with $N'$ denoting the number of rows of $Q_u(H_{ab})$.
	
	We use the (BCH) family of codes as it can correct multiple errors {simultaneously}. Note that $\oplus$ in $(\ref{code})$ is used because the subtraction and the addition in binary fields are the same.
	Bob then recovers $C$ using the recovery process $C' \gets Rec(Q_u(H_{ba}) ,SS(Q_u(H_{ab})))$. From $C'$, Bob decodes $C$ and retrieves $Q_u(H_{ab})$ from $SS(Q_u(H_{ab}))$. We note that hashed value of $Q_u(H_{ab})$ is also broadcasted with $SS(Q_u(H_{ab}))$ such that Bob can confirm whether the correct sequence was recovered. We highlight that this method can guarantee almost zero mismatch, and it is resource and time efficient. After the information reconciliation, both Alice and Bob obtain the same sequence $K$, i.e., $K=Q_u(H_{ab})$ .
	
	
	\subsection{Privacy Amplification}
	We note that attacker can exploit the short-term correlation between adjacent quantized bits in case of high channel sampling or RF switching rate. Messages exchanged during reconciliation can also lead to the information leakage.
	To address this problem, the privacy amplification is needed to generate a high-entropy key from a longer low-entropy secret bit stream. In this work, we utilizes $\textit{leftover hash lemma} $ function to realize the process of privacy amplification.
	
	\section{Proposed PHY-DSSS}\label{sec:phy_dsss}
	
		\begin{figure*}[t]
		\centering
		\includegraphics[width=1\linewidth]{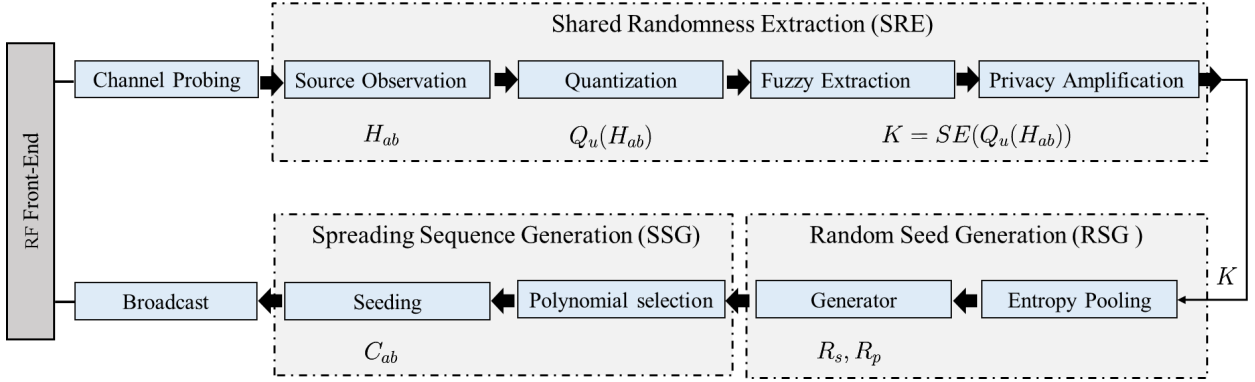}
		
		\caption{Block diagram of the proposed physical layer secret key-assisted DSSS (PHY-DSSS).}
		\label{block}
	\end{figure*}

	In this section, we present the proposed self-controlled jamming resilient scheme,
	which is the physical layer secret key-assisted DSSS (PHY-DSSS).
	The proposed PHY-DSSS utilizes secret keys generated from the physical layer and  DSSS as core elements to enhance the jamming resilience of the adopted system.
	The block diagram of the proposed scheme is provided in Fig. \ref{block}.
	
	In the proposed scheme, Alice and Bob first perform channel probing alternatively and then extract shared secret bits using shared randomness extraction (SRE), which is described in Section~\ref{sec:randomness_ex}. Instead of using the shared secret bit sequence $K$ directly for spreading symbols, we use $K$ to initiate random seeds generation (RSG), which are then used to perform spreading sequence generation (SSG).
	This is because RSG can further remove the spatial-temporal correlation between secret bits generated from consecutive channel estimates. In addition, RSG can generate random seeds based on keys from multiple entropy sources. As such, once RSG is initialized, it can continuously generate random seeds, and consequently reduce the waiting time. Furthermore,  RSG can protect the generated random sequences from being compromised even if the adversary knows the algorithm of RSG \cite{crypto}.

	We highlight that our proposed scheme not only \emph{strengthens conventional DSSS}, but also \emph{relaxes the need for pre-shared sequences}, thereby making it more suitable for massive connected network (e.g., IoT). In DSSS, data signal is spread in the frequency domain using a \textit{chipping sequence} (spreading sequence). As such, the narrow band jamming noise becomes less effective since its power is distributed over the whole band. We denote the ratio  by which interference is suppressed as the processing gain, given by $G = T_{ds}/T_{cs}$, where $T_{ds}$ and $T_{cs}$ are data and chipping sequence period, respectively.
	In the following subsections, we detail the processes of RSG and SSG of PHY-DSSS, and analyze the jamming resilience performance of PHY-DSSS against the broadband jamming attack.

	\begin{figure}[t]
		\centering
		\includegraphics[width=0.4\linewidth]{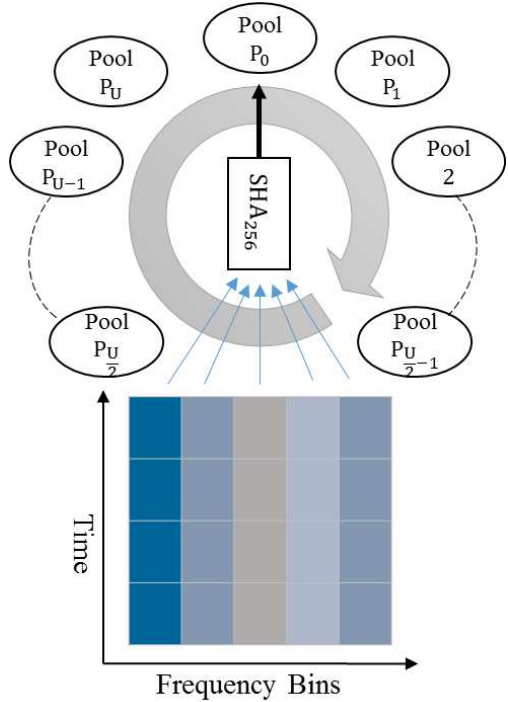}
		\caption{Entropy pooling.}
		\label{pool}
	\end{figure}
	\subsection{Random Seed Generation (RSG)}
	Our RSG is inspired by the random number generator, Fortuna \cite{crypto,eat_entropy}. In particular, we adopt entropy pooling from multiple entropy sources. As such, the internal state of our system is secured. We define input state of RSG as $\mathcal{G} \gets (K,C_p)$ , where $K$ and $C_p$ are the random string generated from SRE and a universal counter, respectively. Both are initialized to 0 at first, which means it has not been fed yet. The details of RSG are as follows:
	
	
	\begin{itemize}
		\item \textbf{Entropy Pooling:} Entropy pooling is designed based on the entropy accumulator in Fortuna using multi-source model. We use OFDM sub-carriers as independent sources of entropy. Each channel estimate matrix is independent from the others and varies with time, location, and frequency \cite{sec_key}. The idea is to increases robustness of the design such that even if any particular spatial location or time slot (due to periodic or random nature of secret bit manipulation by attacker as performed in \cite{Eb}) is compromised, RSG will remain secure until all of observations are compromised.
		
		As shown in Fig. \ref{pool}, channel estimates across different sub-carriers are uniformly distributed over pools, $P_0, P_1,\cdots, P_{U}$, due to the fact that channel estimates are time variant and location independent. Since secret bits keeps refreshing, secrets need to be compressed using SHA-256 hash function to maintain constant pool size of 32 bytes in each pool.
		When pool $P_0$ has accumulated enough randomness, the pool can feed next module, i.e generator can be reseeded.
		\item \textbf{Generator:} The generator, $\mathit{Gen}$, takes a random fixed size string, $h$, from the entropy pool as the input and produces arbitrarily long pseudo-random sequences by using AES-256 as the block cipher \cite{crypto}. As can be seen in Algorithm \ref{rseed}, a counter, $\mathcal{C}_p$, keeps track of number of times the generator has been reseeded from the pools and is used to determine which pool is used to reseed. As is evident from line \ref{line1}-\ref{line2}, pool $P_i$ will be included in the reseed if $2^i$ divides $\mathcal{C}_p$. This insures that pool $P_0$ is used in every reseed, $P_1$ in every second reseed, etc. This means higher numbered pools contribute less frequently but collect a large amount of entropy between reseeding. Reseeding is performed by hashing the specified entropy pools together, using two iterations of $SHA-256$, henceforth denoted $SHA_d-256$. Once a pool has been reseeded, it is then reset to zero. Provided there is at least one source of entropy, OFDM sub-carrier, over which an attacker has no control, it will be unable to predict the contents of at least one pool, and therefore will be unable to break the RSG in this way. The generator operates block cipher on input, $ \mathcal{R}$, in counter encryption mode. In this implementation, $AES-256$ is chosen as the block cipher \cite{crypto}. The plain-text input to the block cipher is simply a counter, $j$, and the 256-bit keys come from the compressed entropy across multiple pools. Since AES is a 128-bit block cipher, it generates 16 bytes of data, random seed $\mathcal{R}_s$ using $\mathcal{R}_s \gets \mathcal{R}_s||E(\mathcal{R}, \mathcal{j})$.	
	\end{itemize}
	\begin{algorithm2e}
		\DontPrintSemicolon
		\KwIn{ $ \mathcal{G} \gets (K, C_p): $ Generator state, $\ s_l$: Seed length\\ }
		\KwOut{$\mathcal{R}_s, \mathcal{R}_p:$ Random seed and Poly-select num}
		\SetKwBlock{Begin}{function}{end function}
		
		\Begin($\text{RSG}$)
		{			
			\If {Update=true}{
				$\mathcal{R}_s \gets \emptyset$\\
				$h \gets \emptyset$	\\
				\For{$ \ i \in 0,...,U$}{
					\If{$ \ 2^i| \mathcal{C}_{p}$}{\label{line1}
						$h \gets h|| SHA_d-256(P_i)$	\\		
						$P_i \gets \emptyset$	\label{line2}		\\	
					}
				}
				$\mathcal{R} \gets SHA_{d-256}(\mathcal{R}||h)$		\\		
				$\mathcal{C}_{p}=\mathcal{C}_{p}+1$
			}
			\For{$j \in 1:round(\frac{s_l}{128})$}{ \label{line3}
				$\mathcal{R}_s \gets Gen(\mathcal{R},j)$ \label{line4}
			}
			$\mathcal{R}_p \gets Gen(\mathcal{G},1)$\\
			$\mathcal{R} \gets Gen(\mathcal{G},2)$				
		}
		\caption{Random Seed Generation}\label{rseed}
	\end{algorithm2e}
	
	Based on Algorithm \ref{rseed}, we find that, along with generating $\mathcal{R}_s$, two more seeds, $\mathcal{R}_p$ and $\mathcal{R}$, are generated as well. We note that $\mathcal{R}$ will be used as the supplementary input for RSG the next time it is required to generate seed. As such, the adversary is not able to estimate the future and past generated seeds even if it has the knowledge of current seed.

	
	\subsection{Spreading Sequence Generation (SSG)}
	We note that $\mathcal{R}_s$ and $\mathcal{R}_p$ are used as the input for SSG. Specifically, $\mathcal{R}_p$ is a 128-bit random binary string, which is used for choosing polynomial from a primitive polynomial bank. The polynomial bank basically consists of lookup table corresponding to possible totatives, \cite{totative}, and combinations of LFSRs. We note that totatives helps define tap location of LFSRs based spreading code generators and randomness in its selection increases complexity of the system, \cite{berk}. Then using the selected polynomial and $\mathcal{R}_s$, LFSR engines generates $2^{s_l} -1$ bits for each $\left(\mathcal{R}_s,\mathcal{R}_p\right)$ at rate $f_s\times r_s$ chips per second, where $f_s$ denotes the spreading factor and $r_s$ denotes the symbol rate at the source.
	
	\subsection{Jamming Resilience Performance Analysis} \label{attack_eq}
	In this subsection, we analyze the performance of the proposed scheme against the broadband jamming attack.
	Specifically, we consider the rate-aware code selection (RACS) attack
	by assuming attacker knows the secret key generation rate, $k_r$.
	%
	As such, attacker can generate all the possible codes for the RSG. For example, if attacker knows that Alice and Bob agree on the secret key of $5$ bits and use this secret key for the RSG, it can then generate all $31$ possible codes. Since attacker does not know the exact code that Alice uses, the best strategy for attacker would be to uniformly allocate its transit power among possible codes and transmit the summation of these codes. As such, the received signal at Bob under RACS attack can be re-expressed as
	\begin{align}
	\label{signal_bob_2} y_b = \sqrt{P_ad_{ab}^{-\alpha}}h_{ab} xC_{ab} +\sqrt{\frac{P_ed_{eb}^{-\alpha}}{S}}h_{eb}x_e\sum_{i=1}^{S}C_{ab,i}+n_b,
	\end{align}
	where $S=2^{k_r}-1$ denotes the number of all possible codes and $C_{ab,i}$ denotes a possible code that attacker generates.
	
	Based on \eqref{signal_bob_2}, we re-express the SINR at Bob under RACS attack as
	\begin{align}
	\label{sinr_bob_2} \gamma_{b} = \frac{\gamma_{ab}|h_{ab}|^2}{\frac{\gamma_{eb}}{S}|h_{eb}|^2\varphi+\frac{1}{L}},
	\end{align}
	where $\varphi=\sum_{i=1}^{S}\varphi_i$, and $\varphi_i$ denotes the correlation between $C_{ab,i}$ and $C_{ab}$. We note that $\varphi_i=1$ if $C_{ab,i}=C_{ab}$.

	In order to evaluate the performance of our proposed scheme under RACS attack, we adopt the successful transmission probability as the performance metric, which is defined as the probability that the received SINR at Bob is larger than a certain threshold. Mathematically, it is given by
	\begin{align}
	\label{performance_metric} P_s\left(k_r,\gamma_{th}\right) = \mbox{Pr}\left(\gamma_b>\gamma_{th}\right).
	\end{align}
	
	According to \eqref{performance_metric}, we now examine the successful transmission probability of our proposed scheme in the following theorem.
	
	\begin{theorem}
		\label{t1} The successful transmission probability of our system under RACS attack is expressed as
		\begin{align}
		\label{t1_result} P_s\left(k_r,\gamma_{th}\right) = \frac{\exp\left(-\frac{\gamma_{th}}{\gamma_{ab}L}\right)}{\frac{\gamma_{th}\gamma_{eb}\varphi}{S\gamma_{ab}}+1}.
		\end{align}
		\begin{proof}
			Based on \eqref{sinr_bob_2}, we derive the successful transmission probability as
			\begin{align}
			&P_s\left(k_r,\gamma_{th}\right)\notag\\ &= \mbox{Pr}\left(\frac{\gamma_{ab}|h_{ab}|^2}{\frac{\gamma_{eb}}{S}|h_{eb}|^2\varphi+\frac{1}{L}}>\gamma_{th}\right)\notag\\
			&=\mbox{Pr}\left(|h_{ab}|^2<\frac{\gamma_{th}}{\gamma_{ab}}\left(\frac{\gamma_{eb}}{S}|h_{eb}|^2\varphi+\frac{1}{L}\right)\right)\notag\\
			&\overset{(a)}{=}\int_{0}^{\infty}\exp\left(-\frac{\gamma_{th}}{\gamma_{ab}}\left(\frac{\gamma_{eb}}{S}x\varphi+\frac{1}{L}\right)\right)\exp\left(-x\right)~\text{d}x\notag\\
			&= \frac{\exp\left(-\frac{\gamma_{th}}{\gamma_{ab}L}\right)}{\frac{\gamma_{th}\gamma_{eb}\varphi}{S\gamma_{ab}}+1},
			\end{align}
			where $(a)$ holds due to the fact that $h_{ab}\sim\exp\left(1\right)$ and $h_{eb}\sim\exp\left(1\right)$. The proof is completed.
		\end{proof}
	\end{theorem}
	We find that Theorem \ref{t1} provides us an efficient tool for evaluating the successful transmission probability of our proposed scheme under RACS attack. The correctness of our analysis in Theorem \ref{t1} will be validated by both simulation and experimental results in Section \ref{sec:results}. We also find that in \eqref{t1_result} the value of $\varphi$ can be approximated by $1+\frac{2^{k_r}-2}{3L}$ using nultiple access interference (MAI) \cite{pursley} and \cite{pursley1}. As such, we obtain an approximation of the successful transmission probability, given by
	\begin{align}
	\label{t1_result_2} \tilde{P}_s\left(k_r,\gamma_{th}\right) = \frac{S\gamma_{ab}\exp\left(-\frac{\gamma_{th}}{\gamma_{ab}L}\right)}{{\gamma_{th}\gamma_{eb}\left(1+\frac{2^{k_r}-2}{3L}\right)}+S\gamma_{ab}}.
	\end{align}

	\subsection{Resilience Against Replay Attack}
	Apart from the RACs attack, we note that our proposed protocol is also applicable for denying the effect of another dominant jamming attack on DSSS systems, i.e., the replay attack. In order to perform the replay attack, attacker first captures signals transmitted by Alice and then re-transmits the captured signals after multiplying them with random symbols. In terms of when the retransmission happens, there can be two attacking strategies for attacker: a) attacking in the same symbol period of Alice transmitting, b) attacking the consecutive symbols. We note that the former attacking strategy is difficult to conduct for attacker, due to the fact that it requires attacker to be located within an ellipse, with focal points at Alice and Bob and the major diameter $d_{ab} + cT_{cs}$, where $c$ denotes the speed of light. However, this requirement is unlikely to be met given the practical limitations such as the processing speed and the precise location awareness \cite{pooper}. The latter attack, on the other hand, does not have such stringent requirements on the processing speed and the location awareness of attacker, and therefore is easier to conduct. Our proposed PHY-DSSS avoids such replay attacks by refreshing codes using RSG and SSG. As such, the cross-correlation between the transmitted signals from Alice and the transmitted signals from attacker is very low.

	\section{Simulation and Experiment Results}\label{sec:results}	
	The proposed randomness extraction and jamming resilient scheme are verified in this section. Specifically, we present simulation results as well as experimental results to validate the performance of our proposed scheme. Since our experimental results are obtained from the testbed in an ideal office environment, we highlight that our proposed scheme is suitable for real-world environments.
	
	
	

	\subsection{Simulation Configuration}
	We use Monte Carlo simulations to show how our proposed scheme can improve the jamming resilience of the considered system. The configurations for Monte Carlo simulations are as follows:
	\begin{enumerate}
		\item Alice, Bob, and attacker are located at (0, 0), (0, 20) and (10$\sqrt{2}$, 10$\sqrt{2}$) respectively.The transmit power at Alice, Bob, and attacker is 45 dBm.
		\item We use OFDM subcarriers to generate secret keys. Following the IEEE 802.11 protocol \cite{sub_wifi}, every 64 subcarriers corresponding to a common pilot OFDM symbol are collected and quantized by using the fuzzy extraction detailed in \ref{fuzzy_extraction}. In addition, we check the correlation between secret keys from adjacent subcarriers. If it exceeds $0.25$, we drop one of the secret keys such that the randomness of generated secret keys are guaranteed.
		\item For entropy pooling, we use $12$ entropy pools. We assume that each time only the channel estimates generated from one subcarrier is fed to entropy pools, thus mimicking the worst-case scenario where other channel estimates are all compromised due to adversarial interference.
	\end{enumerate}
	\subsection{Experiment Configurations}
	We also use a SDR-based testbed to examine the performance of our proposed scheme in realistic environments. We note that our experiments are conducted in the indoor environment. In order to avoid possible interference from existing WiFi links, we adopt the frequency as $2.484$ GHz.	

	\subsubsection{Hardware} 
\begin{figure}		\centering
		\includegraphics[width=4.0in]{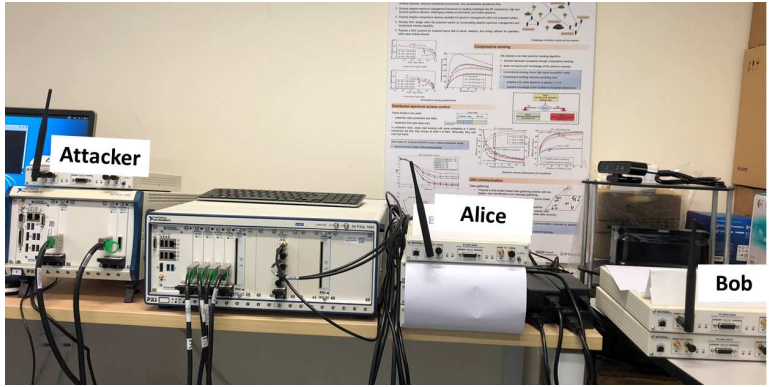}
		\caption{Lab setup and equipments.}
		\label{lab}
	\end{figure}

\begin{figure}
		\centering
		\includegraphics[width=4in]{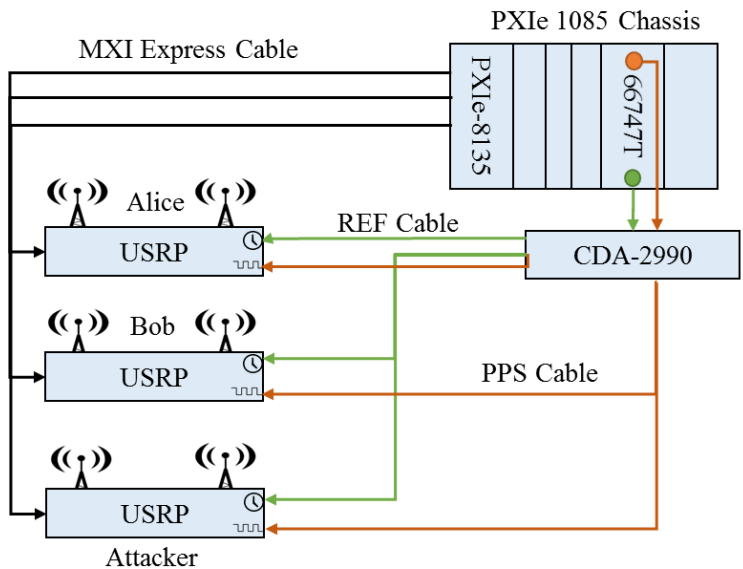}
		\caption{Synchronized USRP.}
		\label{lab_plan}
	\end{figure}

We conduct the experiments in our laboratory with the size of $15~\text{m} \times 20~\text{m}$. As shown in Fig. \ref{lab}, we use Universal Software Radio Peripheral (USRP) 2952 series of National Instruments \cite{ni}. It has 2 full-duplex transmit and receive channels with 40 MHz/channel of real-time bandwidth and a large DSP-oriented Kintex 7 FPGA.
	The analog RF front end interfaces with the large Kintex 7 410T FPGA through dual ADCs and DACs clocked at 120MS/s. Each RF channel includes a switch, allowing for TDD operation on a single antenna using the TX1/RX1 port. In order to realize strict adversary assumptions,  all three USPRs, representing Alice, Bob, and attacker, are interfaced to one PXIe-8135 controller through high speed (800 MB/s) PCI Express x4 cables and PXIe-8384 MXI-Express modules. In addition, we use PXIe-66747T synchronization module to insure transceivers and attacker are tightly synchronized. All modules are stacked in a dedicated chassis, PXIe-1085, as shown in Fig. \ref{lab_plan}. We note that, using the $10$ MHz reference signal and the pulse-per-second singal provided by PXIe-66747T, CDA-2990 clock distribution device can simultaneously drive three USRPs.
	
	%
	%
	\subsubsection{Software and Design}
	We program the USRPs using Labview Communication Design Suite (LCDS) \cite{ni}, which facilitates splitting of the signal processing code blocks between host processor and FPGA. As such, we make sure most intensive and latency sensitive calculations are conducted on FPGA, while the processor only handles control signal and data fetching.
	We implement OFDM symbol based channel estimation FPGA based on LCDS LTE framework \cite{lte}. To guarantee the channel reciprocity, Alice and Bob are required to estimate channels within the coherence time, i.e., 25-30 ms for indoor environments at around 2.484 GHz. Hence, Alice and Bob switch between the transmission mode and the receiving mode fast enough so that they can probe channels within a slotted time. In addition, we consider the scenario where attacker is also capable of fast switching between the transmission mode and the receiving mode in order to inject jamming signals.
	We use automatic transmit/receive (ATRs) on SDR daughter board to achieve the fast switching between the transmission mode and the the receiving mode. The channel probing is operated in $20$ MHz bandwidth mode, which consists of $2048$ frequency-domain subcarriers per OFDM symbol. We organize 1200 usable subcarriers in sets of 12 contiguous subcarriers \cite{ni} and take measurements across each one of them in each TDD slot.

	The entropy pooling and random seed generation are handled by the host controller. Specifically, it first uses seeds to generate spreading sequences, and then sends generated sequences to Alice and Bob for spreading and despreading messages, respectively. Meanwhile, attacker sends broadband jamming signal to interfere with the legitimate transmission.
	
	\subsection{Results} \label{result_channel}

		\begin{figure}
	\centering
	\includegraphics[width=4in]{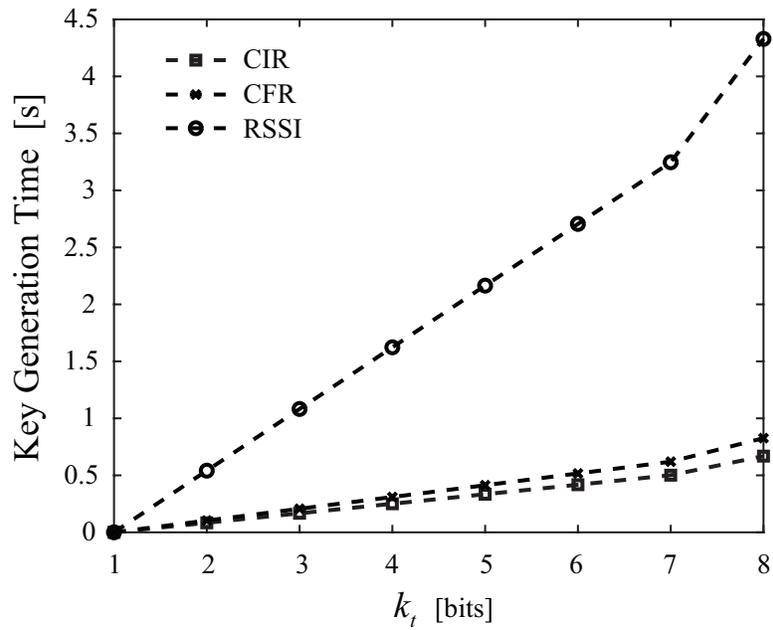}
	\caption{Key generation time versus $k_t$ for different channel measurements.}
	\label{time}
\end{figure}
	
	
	In Fig. \ref{time}, we first examine the impact of the number of secret key bits used per transmission, $k_t$, on keys generation time when secret keys from different types of channel measurements, including RSS, CIR, and CFR, are generated. $k_t$ is used as input to RSG which further drives SSG to generate new codes every symbol.  We note from our experiment that the secret key generation rates of these three types of channel measurements are different. On average, 4, 15 and 16 bits/s can be generated from RSS, CIR and CFR, respectively. We define the key generation time as the time for generating $k_t$ secret key bits from each type of channel measurements. As such, we can see that the key generation time increases as $k_t$ increases. We also see that RSS requires the longest time to generate $k_t$ secret key bits, indicating that RSS can only be employed at system without stringent delay requirements.
	

	%

	\begin{figure}		\centering
		\includegraphics[width=4.2in]{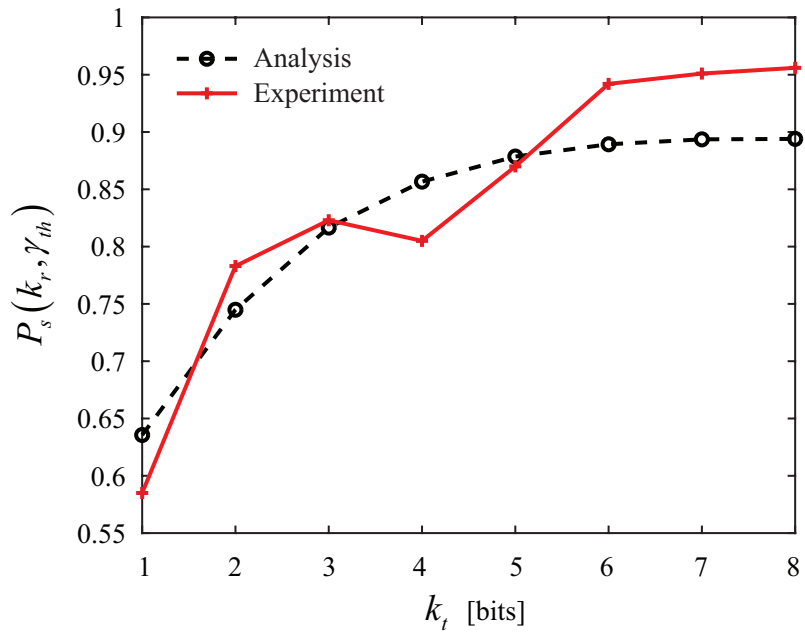}
		\caption{$P_s\left(k_r,\gamma_{th}\right)$ versus $k_t$ in experimental setup and simulation with $\gamma_{th} = 1$ under RACS attack.}
		\label{exp}
	\end{figure}
	\begin{figure}
		\centering
		\includegraphics[width=4in]{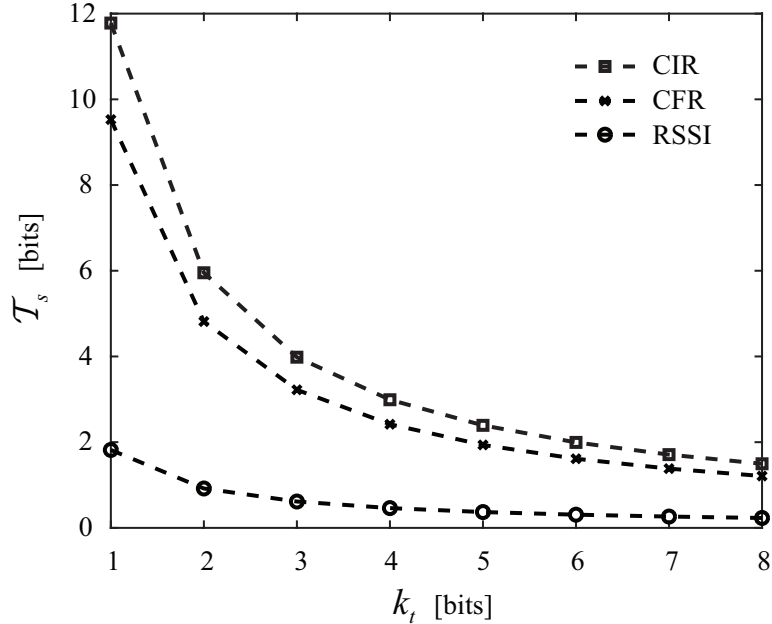}
		\caption{$\mathcal{T}_s$ versus $k_t$ for different channel measurements.}
		\label{succ_t}
	\end{figure}

	In Fig. \ref{exp}, we plot the successful transmission probability versus the number of secret key bits used per transmission, $k_t$, for $\gamma_{th} = 1$. In this figure, we assume that Alice only uses the secret key bits generated from the current block to perform our proposed PHY-DSSS, and drops those secret key bits when the next block comes. The analytical curves and experimental results are marked by black solid lines and red dotted lines, respectively.
	We first see that our analytical curve, generated from Theorem \ref{t1} with $L=1024$, predicts the experimental results, which shows the correctness of our analysis on the successful transmission probability. It also shows that our proposed scheme can improve jamming resilience of systems in real-world environments.
	In addition, we see that the successful transmission probability first monotonically increases in the regime where $k_t$ in relatively small, i.e., $k_t\leq 5$ bits, and then saturates when $k_t$ becomes large. For example, when $L=1024$, the successful transmission probability saturates when $k_t = 7$ bits. This indicates that, even with limited number of secret key bits, the jamming resilience of the system can be supported by using our proposed scheme.
	
	To further understand how well our proposed scheme can improve the jamming resilience of the considered system, we adopt another performance metric, the successful transmission throughput, which is defined as
	\begin{align}
	\label{successful_throughput} \mathcal{T}_s= \frac{k_rP_{s}\left(k_t,\gamma_{th}\right)}{k_t}.
	\end{align}
	Recall that $k_r$ denotes the secret key generation rate, we note that the successful transmission throughput quantifies how many transmission can be successful achieved with a given number of secret key bits. In Fig. \ref{succ_t}, we plot $\mathcal{T}_s$ versus $k_t$ for different lengths of $C_{ab}$ and different types of channel measurements. We first see that $\mathcal{T}_s$ decreases as $k_t$ increases. We then see that $\mathcal{T}_s$ increases as $L$ increases. Moreover, we see that RSS achieves the lowest $\mathcal{T}_s$ in three types of channel measurements.

	\section{Conclusions}
	We proposed a jamming resilient transmission scheme for three-node communication systems, where Alice transmits to Bob in the presence of an attacker. In our proposed scheme, random secret keys were first generated from wireless fading channels. Utilizing generated secret keys to drive cross-layer security design, we show how the jamming resilience of the considered system can be boosted with the aid of DSSS. Using the successful transmission probability as the performance metric, we analytically and experimentally confirm that our proposed scheme can achieve successful transmission in real-world environments even when attacker has the knowledge of the secret key generation rate.
	
	%
	%
	\section*{Acknowledgment}
	The authors would like to thank Dr. Richard Hsu, Dr. Nils Ole Tippenhauer, Ravi Kumar and Dr. Leonid Reyzin for valuable insights during discussions. We are also very thankful to Archit Goyal for sharing java-version codes of Fuzzy extractor.
	\bibliographystyle{IEEEtran}
	\bibliography{ref}
	%
	\IEEEpeerreviewmaketitle
	
	\ifCLASSOPTIONcaptionsoff
	\newpage
	\fi
	%
\end{document}